\documentclass[journal]{IEEEtran}

\usepackage{ifpdf}

\usepackage{cite}
\ifCLASSINFOpdf
 \usepackage[pdftex]{graphicx}
\else
  \usepackage[dvips]{graphicx}
\fi
\usepackage[cmex10]{amsmath}
\usepackage{algorithmic}

\usepackage{array}

\ifCLASSOPTIONcompsoc
 \usepackage[caption=false,font=normalsize,labelfont=sf,textfont=sf]{subfig}
\else
 \usepackage[caption=false,font=footnotesize]{subfig}
\fi
\usepackage{fixltx2e}

\usepackage{stfloats}
\usepackage{fourier} 
\usepackage{array}
\usepackage{makecell}

\usepackage{url}

\begin{document}

\title{Design of nano-scale high static performance phase-change on-off silicon photonic switch}

\author{Nadir Ali\IEEEmembership{}
         and Rajesh Kumar \IEEEmembership{}
\thanks{The authors are with the Department
of Physics, Indian Institute of Technology Roorkee, Roorkee, 247667 INDIA e-mail: (rajeshfph@iitr.ac.in; nadir.dph2016@iitr.ac.in).}
\thanks{}}

\markboth{}%
{Shell \MakeLowercase{\textit{}}}

\maketitle

\begin{abstract}
We report design of a high static performance on-off optical switch using nanoscale phase change material $Ge_{2}Sb_{2}Te_{5}$ embedded into silicon-on-insulator waveguide. This active material can be switched between amorphous and crystalline states using electrical/optical pulses. The fundamental mode propagating through Silicon waveguide drastically alters its properties due to high index change at \bf{$Si-Ge_{2}Sb_{2}Te_{5}$} interfaces and absorption in $Ge_{2}Sb_{2}Te_{5}$. The optical switch made of Silicon waveguide with $Ge_{2}Sb_{2}Te_{5}$ of volume 400 nm $\times$ 180 nm $\times$ 450 nm (length $\times$ height $\times$ width) embedded in to it, provides high extinction ratio of 43 dB with low insertion loss of 2.76 dB in ON state at communication wavelength of 1550 nm. There is trade-off between insertion loss and extinction ratio. For 10 dB extinction ratio, The insertion loss can be as low as 1.2 dB and corresponding active volume is also pretty low (100 nm $\times$ 150 nm $\times$ 450 nm (length $\times$ height $\times$ width)). Further, spectral response investigations reveal that this switch maintains extinction ratio more than 30 dB in wavelength span 1500-1600 nm. The high static performance of optical switch reported here is a direct result of proper dimensional engineering of active volume as well as its incorporation into the silicon-on-insulator waveguide. We also propose figure-of-merit for non volatile optical switches that takes into account relevant parameters of static performance.

\end{abstract}


\IEEEpeerreviewmaketitle

\section{Introduction}

\IEEEPARstart{I}{}ntregated photonic circuits are the way forward for the futuristic communication systems that have potential to overcome the limitations electronic communication systems are facing today, such as latency between processor and memory (Von Neumann bottle neck), high power consumption, and limited bandwidth \cite{sun2015single:1}. Conventional copper interconnects have limitations of loss, speed and cross talk as the density of interconnects increases \cite{miller2009device:2}. Silicon photonics platform is the leading candidate for inter- and intra-chip communication primarily due to low cost and high yield fabrication availability since it can take advantage of highly developed CMOS manufacturing process used in silicon (Si) microelectronics industry \cite{jalali2006silicon:3}.\\
Silicon photonic switches are in great demand to realize Si based photonic integrated circuits as they offer high switching performance like high modulation and picosecond switching time \cite{pelc2014picosecond:4}. For miniaturization of switching devices with high performance, researchers utilized resonant geometries such as ring resonator and photonic crystals. But these devices have low fabrication tolerances, operate in a narrow spectral bandwidth range, and require active temperature control due to high temperature sensitivity \cite{reed2010silicon:5}. Further reduction in footprint can only be done by reducing active ( switching ) volume and at the same time it is also desirable to have high extinction ratio (ER), low power consumption, low insertion loss (IL), and suitability for broadband operation.\\ 
\begin{figure}[!t]
\centering
\includegraphics[width=3in]{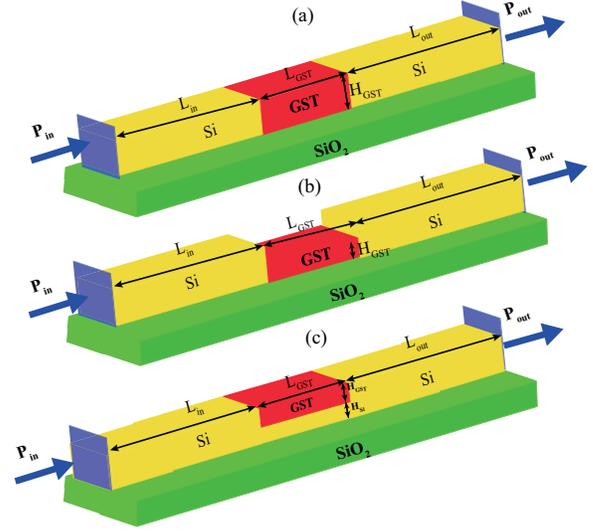}
\caption{Schematic view of switch (a) fully etched Si waveguide with GST cross section matching Si cross section, (b) fully etched Si waveguide with GST height variant, and (c) partially etched Si waveguide and partially filled with GST.}
\label{fig:1}
\end{figure}
Recently, phase transition process has been utilized in optically active chalcogenide materials to realize integrated optical switches \cite{stegmaier2017nonvolatile:6}. Due to their nonvolatile nature, phase change materials (PCM) are suitable candidate for on-chip optical devices that will consume lesser power. PCMs like $Ge_{2}Sb_{2}Te_{5}$ can be switched rapidly and repeatedly between amorphous and crystalline phases by the application of short electrical/optical pulses \cite{hosseini2014optoelectronic:7,kolobov2004understanding:8}. PCMs have several prominent properties such as high optical contrast between amorphous and crystalline state, phase transition time of sub-nanoseconds \cite{wang2008fast:9}, compatibility with CMOS manufacturing process \cite{hwang2003full:10}, and extreme Scalability \cite{lee2007highly:11}, which make PCMs ideal candidate for  chip scale active photonic devices.\\
In the work reported here, a phase change material $Ge_{2}Sb_{2}Te_{5}$ (commonly known as GST) is embedded into a Si strip waveguide in different configurations to realize an ultra-small non-resonant and high static performance on-off switch. In the research work reported earlier \cite{stegmaier2017nonvolatile:6,tanaka2012ultra:12,rude2013optical:13} GST was placed on top of Si waveguide and weak evanescent field coupling was exploited to induce phase transition. Such an approach results in larger switching power and larger active volume. In the approach taken by us, maximum interaction of light with extremely low active volume happens. This paves the way for low power consuming optical switches that can be built using GST for application in Si photonics integrated circuits.
\section{Design and Simulations}
For designing a GST based photonic switch, silicon-on-insulator (SOI) platform is chosen. The Si waveguide is single mode at wavelength of 1550 nm and has a cross-section of $450 \times 220$ $nm^2$, and rests on 2 $\mu$m thick $SiO_{2}$ substrate. Figure \ref{fig:1} shows the schematic views of the proposed optical switch, where phase change material GST in three different ways is incorporated into Si strip waveguide. To incorporate GST, Si is removed either fully or partially in the middle of the waveguide. In the schematic diagrams, $H_{GST}$ and $H_{Si}$ denote height of GST and Si, respectively; in the etched part of the waveguide. The length of GST is denoted by $L_{GST}$ while waveguide length before and after etched part is denoted by $L_{in}$ and $L_{out}$ respectively. In first configuration, Fig. \ref{fig:1}(a), fully etched Si waveguide is filled with GST having the same cross-section as that of Si waveguide and GST length is varied. In second configuration, Fig. \ref{fig:1}(b), fully etched Si waveguide is filled with GST having the same width as Si waveguide and GST height and length is varied. In third configuration, Fig. \ref{fig:1}(c), partially etched Si waveguide is filled with GST having the same width as Si waveguide; GST and etched Si heights are varied in complementary way such that total height ($H_{GST} + H_{Si}$) remains 220 nm. For example, when $H_{Si}$ is 10 nm then $H_{GST}$ is 210 nm and vice-versa. Simulations are performed using optical module of CST Microwave Studio. This software is based on finite integration method, and perfectly matched layer boundary conditions are taken for simulations. The optical properties of device are modeled taking the refractive index of $SiO_{2}$, Si and air as 1.45, 3.48 and 1, respectively at 1550 nm wavelength. For GST, we use the value of refractive index $n_{am} + \iota \kappa_{am} = 4.67 +\iota 0.12$ for amorphous phase and $n_{cr} + \iota \kappa_{cr} = 7.25 +\iota 1.49$ for crystalline phase at 1550 nm wavelength \cite{shportko2008resonant:14}, where $n$ and $k$ denote real and imaginary part of refractive index respectively.\\ 
The phase change material GST embedded into Si strip waveguide affects the propagating mode depending upon the phase of this material. The light launched from input port traverse $L_{in}$ length in Si waveguide then interacts with GST, and traverse $L_{out}$ in the Si waveguide before reaching output port. In crystalline phase, due to the large value of extinction coefficient, GST behaves as a lossy waveguide and propagating mode suffers large attenuation as most of the light gets absorbed into it. In contrast, mode propagates with lesser attenuation in the amorphous phase due to smaller 
\begin{figure}[!t]
\centering
\includegraphics[width=3.5in]{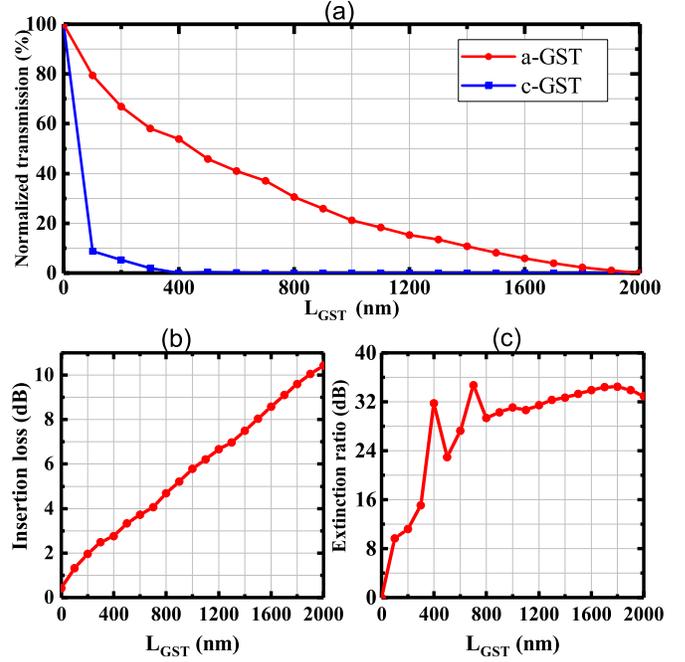}
\caption{(a) Normalized transmission through GST incorporated Si waveguide as a function of $L_{GST}$ with a-GST (red line) and c-GST (blue line), (b) insertion loss and (c) extinction ratio as a function of $L_{GST}$.}
\label{fig:2}
\end{figure}
value of extinction coefficient as compared to that of crystalline phase. Therefore, amorphous and crystalline phases of GST corresponds to ON and OFF state of the switch, respectively. The dimensions of GST are optimized in order to achieve acceptable IL, high ER and ultra low footprint. In order to find best possible configuration of the switch following steps are followed. First, transmission is calculated for all three cases by dimensional variation as described earlier in this section. The transmission data obtained, for each case, for amorphous and crystalline phase of GST is used for calculation of ER $ = 10 \log({T_{ON}}{/T_{OFF}})$, where $T_{ON}$ and $T_{OFF}$ denotes transmission corresponding to amorphous and crystalline phase of GST, respectively. Same transmission data is used for calculation of IL. In our case IL is defined as $ 10 \log T_{ON}$. For the configuration that gives highest ER and suitable for fabrication, spectral response is calculated in wavelength span of 100 nm around central wavelength of 1550 nm. 
\section{Results}
To numerically investigate the dependence of switch parameters IL and ER on the dimensions of GST film, simulations are performed for both amorphous and crystalline phase for various physical configuration of switch. Results of the first set of simulations that correspond to the Fig. \ref{fig:1}(a) are shown in Fig. \ref{fig:2}. In this case embedded GST length $L_{GST}$ is varied from 0 to 2 $\mu$m along the direction of mode propagation at the intervals of 100 nm while keeping cross-section same as that of Si waveguide. When light propagating in Si waveguide reaches etched part filled with GST, losses occur due to absorption in GST and reflections at Si-GST interfaces. These losses are higher in case of crystalline GST (c-GST) as compared to those in amorphous GST (a-GST). This is due to imaginary as well as real part being much higher for c-GST than those for a-GST. Consequently, transmission through waveguide decreases exponentially as a function of GST length in case of a-GST. However, due to highly lossy nature of crystalline phase, transmission decreases much sharply as length increases. The percentage transmission for both the phases is plotted in Fig. \ref{fig:2}(a), and values of IL and ER are shown in Fig. \ref{fig:2}(b) and \ref{fig:2}(c), respectively. 
\begin{figure}[!t]
\centering
\includegraphics[width=3.5in]{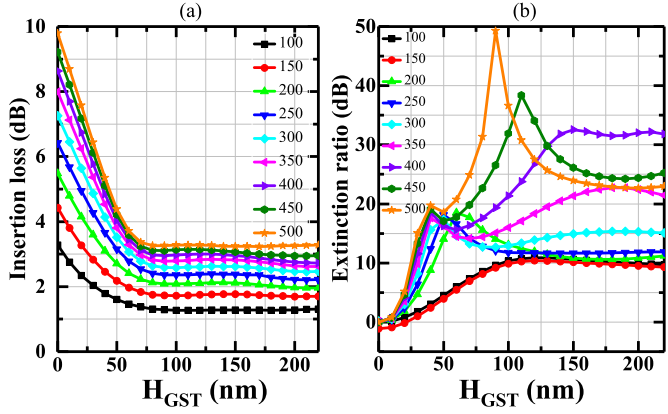}
\caption{(a) Insertion loss and (b) extinction ratio for fully etched GST embedded Si waveguide with $H_{GST}$ variation at different values of $L_{GST}$.}
\label{fig:3}
\end{figure}
We observed highest ER of 34.73 dB at $L_{GST} = 700$ nm at the expense of quite high IL of 4.3 dB. Such a high value of IL is undesirable when it comes to minimize overall power consumption of the device. In the second set, Si waveguide is completely etched and GST is embedded in the channel with variation of height $H_{GST}$ [Fig.\ref{fig:1}(b)] from 0 to 220 nm at different values of $L_{GST}$ from 100 to 500 nm. Results of the second set of simulations are shown in Fig. \ref{fig:3}.These simulations clearly exhibit dependence of ER and IL on the GST height. It may be seen that IL decrease steadily for the value of $H_{GST} < 80$ nm and then become constant for each value of $L_{GST}$. This saturation of IL is due to the mode supported at height 80 nm and above in the GST film. Figure \ref{fig:3}(b) illustrates ER trend. The maximum ER of approximately 49.22 dB is obtained at $L_{GST} =$ 500 nm and $H_{GST} =$ 90 nm with IL of 3.25 dB. This configuration of switch with GST height 90 nm gives the highest ER, but deposition of metal electrode on top of GST will result in large loss in amorphous phase because mode propagating through the waveguide will pass through the metal electrode. Subsequently IL will considerably increase in amorphous phase. To remove these limitations we partially etched Si waveguide and filled empty space with GST as shown in Fig. \ref{fig:1}(c). In this case, parameter sweep was performed by varying heights $H_{GST}$ and $H_{Si}$ from 0 to 220 nm at different values of $L_{GST}$, and $H_{GST}$ is complementary with $H_{Si}$ and vice-versa. Results for this set of simulations are shown in Fig. \ref{fig:4}. Results in Fig. \ref{fig:4}(b) show ER trend similar to that of partial etch, and maximum ER of 43 dB is observed at $H_{GST}$ = 180 nm and $L_{GST}$ = 400 nm with IL of 2.76 dB.\\
The optical switches designed here are of non volatile nature and for their static performance relevant parameters are ER, IL and active volume. The electrical bias is not relevant here. Therefore, in order to have a comprehensive idea of designed optical switches. We propose figure-of-merit (FOM) defined as FOM $= ER / (IL \times active $ $volume)$, where ER and IL values are taken as linear. The FOM values corresponding to highest ER of each configuration are listed in Table \ref{table:1}. At last, spectral response of ER is calculated for optimized geometry of switch, that corresponds to Fig. \ref{fig:1}(c) to find its suitability for broadband operation. As shown in Fig. \ref{fig:5}, the values of ER more than 30 dB are obtained for wavelength span of 100 nm around communication wavelength 1550 nm.\\ 
\begin{figure}[!t]
\centering
\includegraphics[width=3.5in]{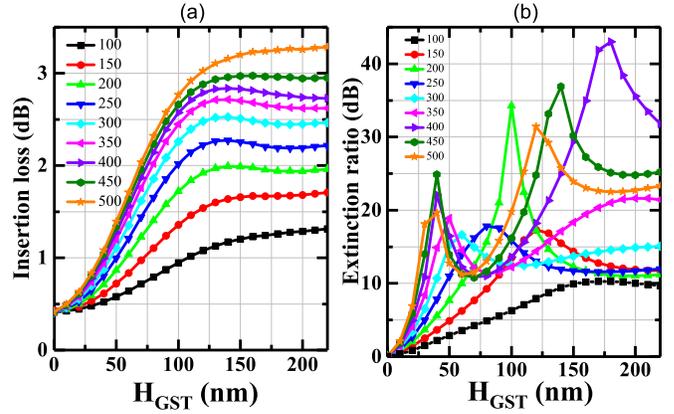}
\caption{(a) Insertion loss and (b) extinction ratio for partially etched GST embedded Si waveguide with $H_{GST}$ variation at different values of $L_{GST}$.}
\label{fig:4}
\end{figure}
\renewcommand\theadalign{bc}
\renewcommand\theadfont{\bfseries}
\renewcommand\theadgape{\Gape[4pt]}
\renewcommand\cellgape{\Gape[4pt]}
\begin{table}[!t]
\caption{FOM values corresponding to highest ER.}
\label{table:1}
\centering
\begin{tabular}{|c|c|c|c|c|}
\hline
\thead{Design} & \thead{Highest ER \\ (linear)} & \thead{IL \\ (linear)}& \thead{Active volume \\ ($ \times 10^{-20} m^{3}$)} & \thead{FOM \\ $\times 10^{22}m^{-3}$} \\
\hline
Fig. 1(a) & 2971.66 & 2.69 & 6.93 & 1.59 \\
Fig. 1(b) & 97723.72 & 2.11 & 2.02 & 229.28 \\
Fig. 1(c) & 19952.62 & 1.82 & 3.24 & 34.02 \\
\hline
\end{tabular}
\end{table}
\begin{figure}[!b]
\centering
\includegraphics[width=3.5in]{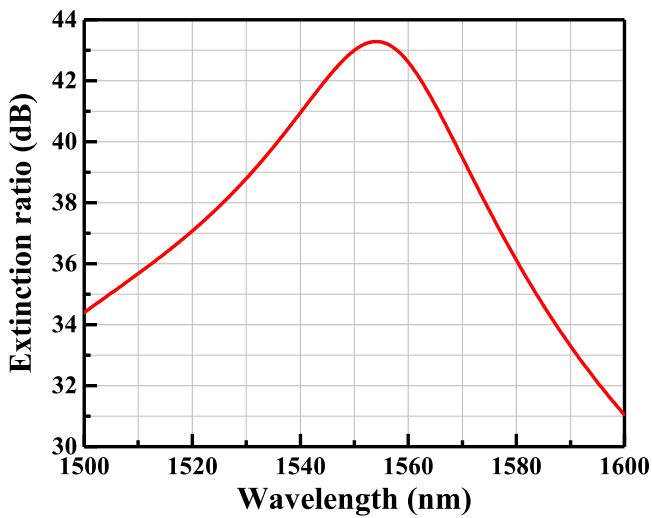}
\caption{Spectral response of extinction ratio for optimized geometry GST on partially etched Si waveguide switch in the wavelength range of 1500 to 1600 nm.}
\label{fig:5}
\end{figure}

\begin{figure}[!t]
\centering
\includegraphics[width=3.5in]{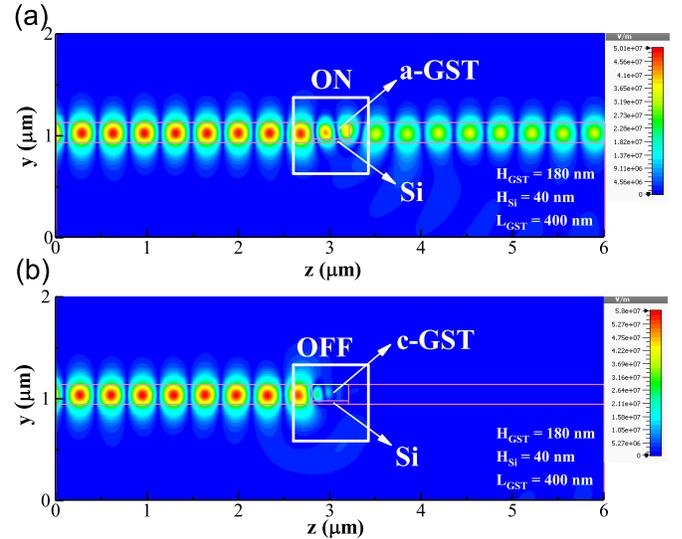}
\caption{Illustrations of electric field profile for optimized switch geometry with dimensions $H_{GST}$ = 180 nm, $H_{Si}$ = 40 nm, and $L_{GST}$ = 400 nm for (a) ON state (a-GST) and (b) OFF state (c-GST).}
\label{fig:6}
\end{figure}

To demonstrate the working of designed optical switch, absolute value of electric field profile is plotted for optimized geometry of GST on partially etched Si waveguide and same is shown in Fig. \ref{fig:6}. The refractive index contrast between amorphous and crystalline phases of GST results in different electric field profile of propagating mode and therefore different output power at the end of waveguide. In amorphous phase Fig. \ref{fig:6}(a), mode propagate through the GST without any appreciable attenuation due to smaller value of extinction coefficient and thus with high transmission depicts ON state of the switch. In contrast, crystalline phase is highly lossy and the mode traveling through the GST suffers larger attenuation due to high absorption. Therefore, with almost negligible transmission, crystalline phase depicts the OFF state of the switch as shown in Fig. \ref{fig:6}(b).
\section{Discussion and Conclusion}
In summary, we designed and numerically investigated an ultra-compact, non-resonant and hybrid Si-GST integrated optical switch for high static performance. We utilized three different approaches for optimization of embedded GST to obtain high ER and low IL. For each case, FOM values are calculated corresponding to highest ER. If we consider losses incurred due to the metal electrodes deposited on the GST, the mode propagating through waveguide will suffer attenuation depending upon the placement of metal electrodes. Subsequently, the value of IL and ER will change. So, electrodes have to be deposited in such a way that the loss is minimum especially in case of the amorphous phase. In first approach $H_{GST}$, and in third approach optimized $H_{GST}$ and $H_{Si}$ put together, are equal to total height of Si waveguide. Therefore, electrodes deposited on GST with a buffer layer will not affect the propagating mode. However, in second approach the optimized GST height is 90 nm and a metal electrode deposited on top of GST will attenuate the propagating mode.\\
The results of this study reveal that variation in dimensions of GST embedded  into Si waveguide leads to variation in performance of optical switch. Partially etched Si waveguide embedded with GST with size 400 nm $\times$ 180 nm $\times$ 450 nm (length $\times$ height $\times$ width) could offer switch performance with high ER of 43 dB and low IL of 2.76 dB at communication wavelength 1550 nm. The simulation results show the trade-off between ER and IL. If we consider the  ER of 10 dB, the IL can be as low as 1.2 dB and corresponding active volume is also very low. Furthermore, the spectral response of this optimized switch geometry shows ER of more than 30 dB for a wavelength span of 100 nm around communication wavelength of 1550 nm. This work lays foundation for realization of ultra-low footprint, high ER, low IL and low power consuming PCM based hybrid integrated electro-optic switches, photonic memories and electro-optic modulators on silicon platform for inter- and intra-chip communication applications.

\section*{Acknowledgment}
Authors sincerely acknowledge the funding received from IIT Roorkee (FIG scheme); and Science and Engineering Research Board (SERB), DST, Govt. of India under Early Career Research award scheme (File no. ECR/2016/001350).
\ifCLASSOPTIONcaptionsoff
  \newpage
\fi
\bibliographystyle{IEEEtran}

\bibliography{sample}


%










\end{document}